# Security, Privacy and Trust: Cognitive Internet of Vehicles

Khondokar Fida Hasan[1], Anthony Overall[1], Keyvan Ansari[2], Gowri Ramachandran[1], Raja Jurdak[1]

1 School of Computer Science, Queensland University of Technology (QUT)
2 School of Science, Technology and Engineering, University of the Sunshine Coast (USC)

## I. Abstract:

The recent advancement of cloud technology offers unparallel strength to support intelligent computations and advanced services to assist with automated decisions to improve road transportation safety and comfort. Besides, the rise of machine intelligence propels the technological evolution of transportation systems one step further and leads to a new framework known as Cognitive Internet of Vehicles (C-IoV). The redefined cognitive technology in this framework promises significant enhancements and optimized network capacities compared with its predecessor framework, the Internet of Vehicles (IoV). CIoV offers additional security measures and introduces security and privacy concerns, such as evasion attacks, additional threats of data poisoning, and learning errors, which may likely lead to system failure and road user fatalities. Similar to many other public enterprise systems, transportation has a significant impact on the population. Therefore, it is crucial to understand the evolution and equally essential to identify potential security vulnerabilities and issues to offer mitigation towards success. This chapter offers discussions framing answers to the following two questions, 1) how and in what ways the penetration of the latest technologies are reshaping the transportation system? 2) whether the evolved system is capable of addressing the concerns of cybersecurity? This chapter, therefore, starts presenting the evolution of the transportation system followed by a quick overview of the evolved CIoV, highlighting the evolved cognitive design. Later it presents how a cognitive engine can overcome legacy security concerns and also be subjected to further potential security, privacy, and trust issues that this cloud-based evolved transportation system may encounter.

## II. Introduction:

The transportation system is an indispensable part of modern civilization and a primary contributory sector in today's economy. In the last two decades, the vehicular industry has experienced sharp growth, especially in first world countries. According to the Motor Vehicle Census in 2020, there are 19.8 million registered motors for 25 million people in Australia, with an annual average increase of 1.7% [2]. The whole world has more than 1.4 billion vehicles, and this number is projected to be doubled in 2040 [3]. With an increase in cars on the roads, there comes a substantial rise in traffic-related incidents, such as vehicle and pedestrian accidents and increased congestion. However, noticeable efforts in developing the transportation system are observed so far. Among them, the recent technological development utilizing cloud-based artificial intelligence and machine learning towards developing cognitive computing technology is considered a revolutionary step in addressing some burning issues such as driving behavior and providing the necessary rapid response on the road.

Meanwhile, the tech world experienced penetration of the novel concept of communication between humans and things and among things themselves over the last few decades. The Internet of Things (IoT), as it is named, is a technological advancement that enabling such interactions mobilizing the Fourth Industrial Revolution. The concept of IoT is to make every single 'network enabled' object in the world connected and represents a vision in which the internet extends into the real world, embracing everyday objects from the Internet of Computers to the Internet of Things. This paradigm already demonstrated its potential and reshaping the future of communication, bringing further improvements and radical transformations to human lives, including homes, transportation systems, the environment, and human well-being [4].

Under the umbrella of the Internet of Things (IoT), transportation evolution is considered a breakthrough in respect of trends and traffic management approaches to enable safety and comfort on the road is also popularly termed as the Internet of Vehicles (IoV). IoV makes sensor platforms that can receive information from other vehicles, the environment, and the driver to ensure a safer road transport system.

Although a significant improvement in automation and connectivity is observed, IoV based model is still inefficient due to the lack of technical sophistication to reduce the road causalities to zero [5]. The IoV framework-based solution cannot address many issues. One of the primary issues, perhaps, is the driving errors and misjudgment by drivers. For autonomous vehicles, this can relate to onboard sensors' errors and misinterpretation of data. According to road research statistics, over 90% of road accidents are caused by humans at present [6]. Such accidents are primarily due to fatigue while driving, overspeeding, blocked line of sight in the road are directly caused by human factors such as cognitive limitations and judgments. Limits of human and existing technology-based solutions encourage the necessity of applying emerging technologies such as machine-enabled cognition using Machine Learning (ML), Artificial Intelligence (AI), and related technologies with system automation that can control decision-making. It can offer technical support to enable error-free driving, emergency response and advanced driving assistance resulting in the idea of the Cognitive Internet of Vehicles (CIoV). CIoV is an evolved paradigm of IoV that introduces cloud-based AI/ML into the transportation system. Therefore, cognitive technology has allowed for more significant enhancements in IoV and its capabilities. However, cybersecurity issues pose the greatest threat to effective implementation of the future to safe mobility.

In the following sections, at first, a brief of intelligent transportation systems' evolution is presented. Later a close look at the evolved framework, the Cognitive Internet of Vehicles (C-IoV), is discussed before outlining the primary issues of Security, Privacy, and Trust around the developed layer in the cloud.

### III. EVOLUTION OF INTELLIGENT TRANSPORTATION SYSTEM

Transportation systems constitute an indispensable part of modern life that plays a critical role in coordinating all forms of transport and related traffic to offer safety by tackling a range of challenging issues. Since the invention of the wheel, we have witnessed the evolution of transportation in all forms, from automobiles to personal commuters like bicycles, motorcycles, scooters, etc. This evolution was perhaps slow over thousands of centuries, has gained impetus in the last two centuries. However, the greatest breakthrough is observed perhaps in the last two decades with the introduction of Information and Communication Technologies (ICT) to the transportation sector. ICT is the driving force behind some of the remarkable innovations in the transportation industry in modern society. Novel ICT-enabled technologies are irresistibly integrating into automobiles to improve driving experiences and tackle some burning issues, such as traffic congestion, road accident, and other road fatalities that causes death and property lost.

Aiming towards smart transport and traffic management systems, combining different technologies and different modes of applications, in 19$^{th}$ century, the umbrella term Intelligent Transportation System (ITS) has emerged and is widely used to identify transportation innovation. In 20$^{th}$ century, however, we had observed remarkable innovation and rapid penetration of disruptive ICT technologies over the last twenty years. This leads to having three successive frameworks after ITS are: Cooperative ITS (C-ITS), Internet of Vehicles (IoV), and Cognitive Internet of Vehicles (C-IoV).

A graph showing the evolution of intelligent transportation is presented in Figure: 1, outlining the primary technical features, different stages of the cloud and market inception of the prime technology. Four significant phases from the 1980s can be seen where around 2010, Autonomous cars, VANET, and LTE networks were incepted into the market. With CITS, both vehicular cloud and edge cloud are introduced. At the same time, the Internet cloud is the platform of the frameworks IoV and CIoV.

While defining, ITS is a set of technologies, applications, and services to tackle road safety issues and empower mobility to assist with productivity and comfort. Conceptually, the term ITS broadly refers to a transportation system with various sensors and assisting technologies integrated on vehicles and infrastructures to monitor their local environment to understand their surroundings accurately. Additionally, wireless communication lets vehicles exchange information with each other and with other road elements.

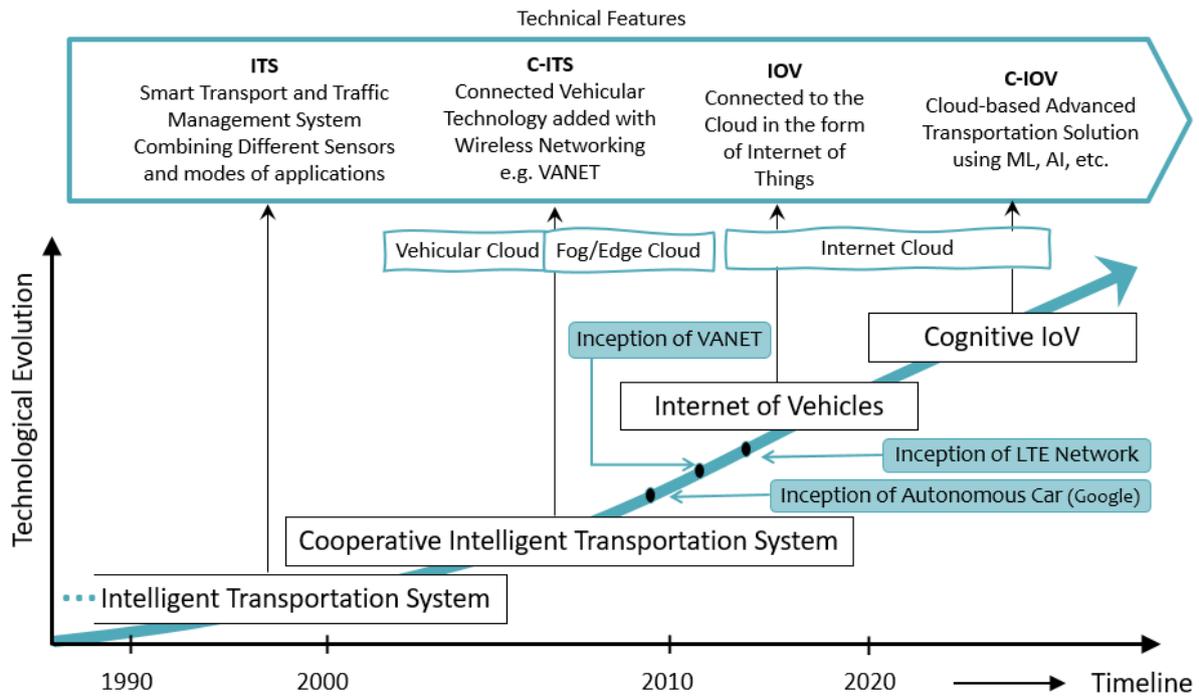

*Figure 1 Evolution of Intelligent Transportation System [7]*

Although both sensor-based autonomous and communication-based assisted technology march in parallel, the technological evolution under each set-up has been treated differently, and different groups of scientists, researchers, and developers are working under two technology directions: Autonomous vehicles (AV) and Connected Vehicles (CV). Thus, technologically, the Intelligent Transportation System generally refers to a range from Autonomous Vehicle (AV) to Connected Vehicles (CV), as shown in Figure 2.

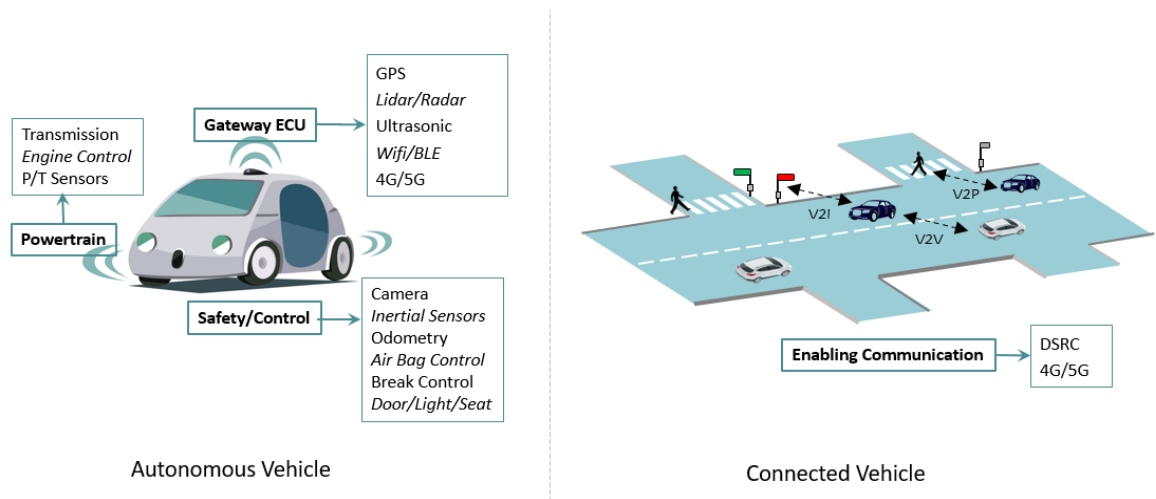

*Figure 2  The concept of Intelligent Transportation System. (a) Autonomous Vehicle (b) Connected Vehicles [1]*

More closely, in autonomous vehicles, a broad range of sensors is integrated into the vehicle to collect the data about the surrounding environment enabling them to operate independently. Primarily, autonomous vehicles can be considered a self-controlled robot that operates independently and takes a wide range of decisions onboard without human interaction [8]. While traveling on the road, these vehicles can perceive, understand and interpret traffic scenarios, make intelligent decisions, and act upon them. The ability to operate independently is facilitated mainly by the innovation in robotics and available tools and techniques. Noticeable breakthroughs in robotics have been observed in recent years due to the advancement of Artificial Intelligence (AI). AI enables machines to make independent, intelligent decisions like humans. AI-powered vision and signal processing techniques gather information around the road environment, interpret and model the information to make necessary decisions. The actions may further change the environment, requiring subsequent decisions. This results in a close-looped system that tightly integrates the physical and digital realms.

A single vehicle presumably has limited perception ability on the road, therefore, limited knowledge about the road and the other vehicles' intent. Thus the Connected Vehicle (CV) emerged to allow the vehicles to communicate with each other and share information about vision, diving intents and other road information such as weather reports, traffic scenarios and even entertainment for the passengers.

In Connected Vehicles (CV), drivers have access to wireless connections to other neighboring vehicles, infrastructure, pedestrians and other devices within their proximity. The principal advantage of having wireless communications is the ability to access information that may otherwise be beyond the driver's immediate awareness. Such wireless communications would help prevent possible collisions by exchanging status information (such as the location, speed and direction of travel of nearby vehicles etc.), and event-driven safety messages (such as lane changing and collision warnings etc.). Along with these safety warning messages, such communication would also support sharing traffic information, weather updates and Internet-based infotainments.

At its earlier stage, this technological endeavor evolved with integrating cooperative and assisting communication technology termed as Co-operative Intelligent Transportation System (CITS), which is essentially enabled by Vehicular Ad-hoc Networks (VANET). Vehicular Ad Hoc Network (VANET) is a special kind of Mobile Ad-Hoc Network (MANET) to the domain of vehicles on the road. The basic communication architecture in VANETs consists of two blocks: On-Board Units (OBU) which is essentially the vehicle, and Road Side Units (RSU).

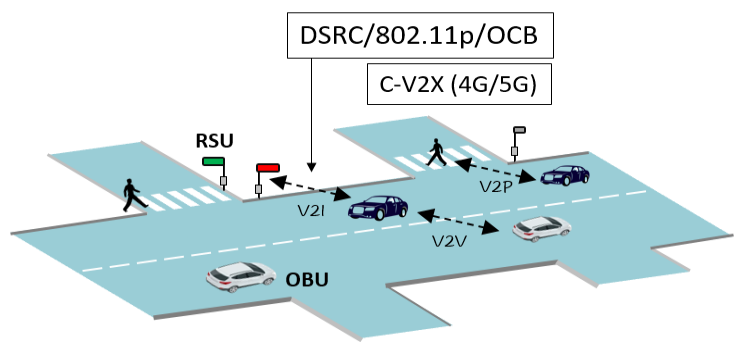

*Figure 3 Vehicular Adhoc Network (VANET)*

The idea of VANET received significant attention in both academia and industry helped to flourish it over the past decade. The core principle of VANET is that a vehicle connects to other vehicles and roadside units to share and propagate information. Meanwhile, with the advancement of the Internet of Things (IoT), vehicles are being connected to the internet. It is aiming at providing ubiquitous access to information alike to the drivers and passengers.

Within this merge, vehicles are in the process of evolution equipped with a range of sensors, powerful onboard computational units, Internet connectivity over IP, a range of communication technologies with direct or indirect connection capabilities. Those evolved features enabled new characteristics of the data traffic in the vehicular network to the concept of Big Data [9].

With this paradigm shift, one of the critical issues VANET faces is its incapacity to process large amounts of data collected by themselves and other devices around them. This overall leads to the technological evolution named Internet of Vehicle (IoV). The IoV framework includes the internet as a permanent asset to the vehicular network systems. It allows forming an interconnected set of vehicles to offer Internet-based services such as storage and computation to assist with road safety, road management, and infotainments [10].

In contrast to VANET, IoV fundamentally has two interrelated but separate technological directions: Internet-based networking and intelligence service. Internet-based networking is primarily enabled through a combination of three; VANET (known as vehicles' interconnection), Vehicle Telematics (known as connected vehicles) and Mobile Internet.

On the other hand, intelligence is assumed in two stages. One is onboard to the vehicle, which combines driver and vehicle as unity using internal onboard intelligent sensors and technologies. The other is cloud-based intelligence. Overall, IoV is a centralization concept compared to VANET that focuses on the intelligent integration of humans, vehicles, things and environments around it in a meaningful way to provide the necessary services on the road.

While conceptualizing the cloud services in the vehicular ecosystem, it can be seen that there are three levels of cloud; vehicular cloud, fog/edge cloud and Internet cloud, as shown in Figure 4. In the vehicular cloud concept, the vehicles on the road communicate with each other and share their computing resources, storage resources, and spectrum resources.

Every vehicle should have access to the cloud to utilize services. The prime motivation behind this paradigm is that; it use to happen, vehicles spend many times in rest, for example, in a parking garage, driveway, parking lot, or even on the road. Since vehicles have resources, the resources can be potentially used in all those situations, especially the parked vehicles that may have vast unemployed and wasted resources. Considering these features, vehicles are considered to form a vehicular cloud network [11]. In comparison to an individual vehicle, the vehicular cloud offers more resources. [12].

Fog or edge cloud services, on the other hand, are hosted in the vicinity of end-users (edge of the network, roadside units). It also includes computation, storage, and networking. The prime motivation of having fog is to complement the centralized cloud by moving down some computing resources to the edge to offer services with reliable access to some delay-sensitive mobile applications [13].

Compared to the other two cloud formation, the Internet cloud is vast, where internet-based services can offer extensive storage, powerful computation, and scalability. Such an Internet cloud can be public, for example, to access infotainment services. It can also be private, for example, cloud services of any car manufacturing company, or it can be an enterprise cloud such as the back office of the road transportation system [10]. Such internet and cloud-centered network is the basis of the concept of Internet of Vehicles (IoV)

With the recent rapid advancement of Internet of Things and future autonomous driving expectations, the Internet of Vehicles (IoV) has attracted wide attention. The high penetration of sensors onboard and roadside generates trillions of data propelling to leverage the cloud-based IoV resources for big data processing.

However, the emerging issues on how IoV could handle the big data intelligently and how the big data-enabled IoV could better support the ITS applications and improve the driving experiences are not well understood with the proposed IoV structure [5]. Additionally, network automation, stable connectivity, and reasonable connectivity and service cost to optimal uses of the resources cannot be guaranteed by the existing proposed IoV framework utilizing cellular networks and ad-hoc networks [5].

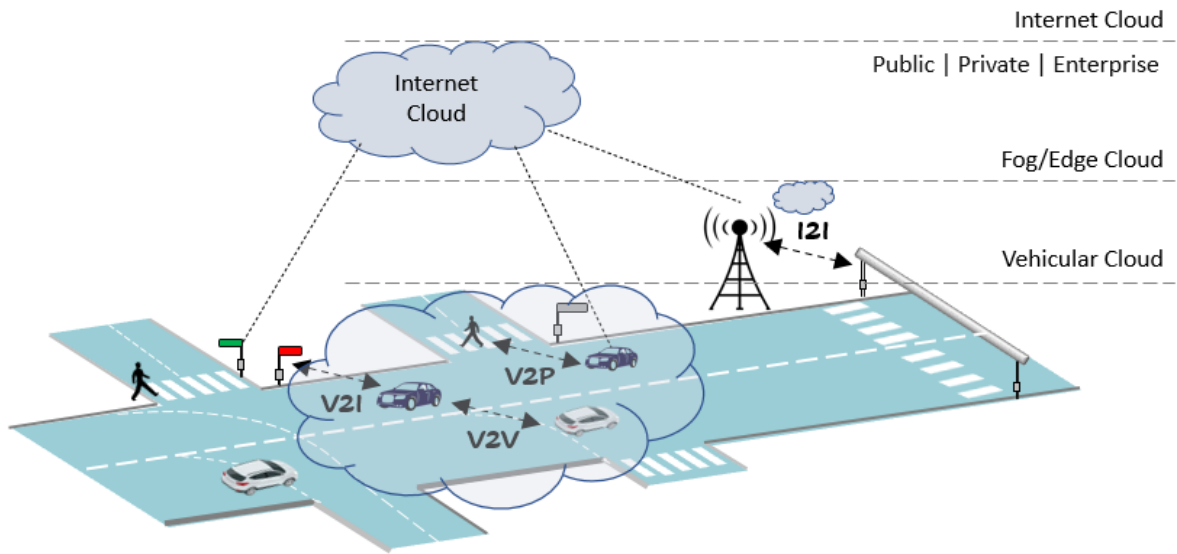

*Figure 4 Conceptual Diagram of Internet of Vehicles.*

To alleviate those issues, a redefined and well-structured framework named Cognitive IoV is first proposed by a group of researchers [5, 14]. This framework utilizes artificial intelligence, cloud/edge computing and 5G network slicing to deal with the top-level issues of IoV, such as the comprehensive modeling of the intelligence access to the network, automation in network control and network healing.

The overall development and relationship are presented in Figure 5, and a close look at the cognitive Internet of Vehicles (C-IoV) architecture is presented in the following section.

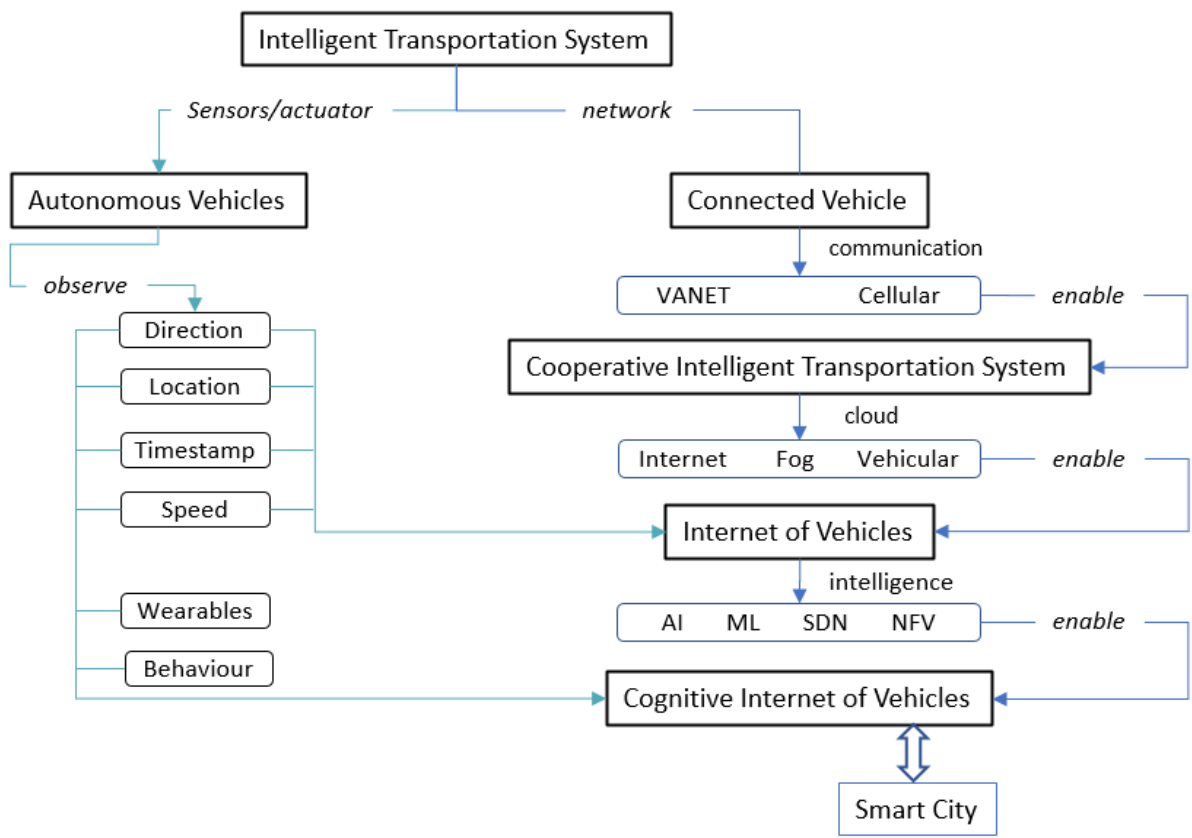

*Figure 5 High-Level Ontology to Map development relationship*

## IV. Cognitive Internet of Vehicles (C-IoV): Motivation and framework

### A. Overview of C-IoV

Road traffic is fundamentally a dynamic network that usually hosts enormous numbers of vehicles and commuters passing through each other and also the road infrastructure at every point in time. Utilizing collected data from different deployed sensors and shared information on the road, it is possible to locate traffic congestions, deadlocks on the road, slow-moving traffic and other road-related affairs by advanced data analytics modeling. In recent days there have been significant attempts to model and study traffic more effectively than ever, thanks to big data analysis. This development attempts to make meaningful operational policies that could maximize resource utilization and lead to a safe travel experience.

The original idea of the Internet of Vehicles (IoV) is limited to the concept of connecting vehicles and road infrastructures to the internet to share their observation and receiving existing internet-based services such as infotainments. However, it is now understood that the idea of only connection over the internet and its limited service is not enough to meet the demand and expectation. In fact, beyond that, the internet cloud can be employed to enable the capability to sense, understand, learn and think independently from both the physical and social worlds by themselves utilizing advanced technologies. Such expectation leads to having a new paradigm named Cognitive Internet of Vehicles (C-IoV), which supports cognitive computing capacity to the network employing Machine Learning, Artificial Intelligence, Software Defined Networking and similar technologies. It fundamentally aims at bridging the transportation system such as a vehicle, road infrastructure and the social world such as human demand, awareness and social behavior and so on. It also aims to enable smart network operation and optimization, resource allocation, emergency responses and intelligent service provisioning [5, 7, 15, 16].

*However, the question is, what does Cognitive refer to here?* The term "cognition" originally received from the general definition of human (animal) cognition system of intelligent behaviour, which refers to the states, experience, and process of knowing by utilizing senses and all conscious and unconscious processes [17]. The process includes perceiving, recognizing, and reasoning for intuitive problem solving and decision-making.

Cognitive in the context of engineering, on the other hand, generally refers to cognitive computing, a branch of computer science that deals with computerized models to simulate the human thought process in complex situations. Using self-awareness and self-learning algorithms and models that use advanced artificial intelligence and techniques such as data mining, pattern recognition, and natural language processing, the computer can mimic the human brain's works. There is indeed much talk about machine learning and artificial intelligence, and recently cognitive computing. The following analogy with a human brain can clarify the differences between these terms.

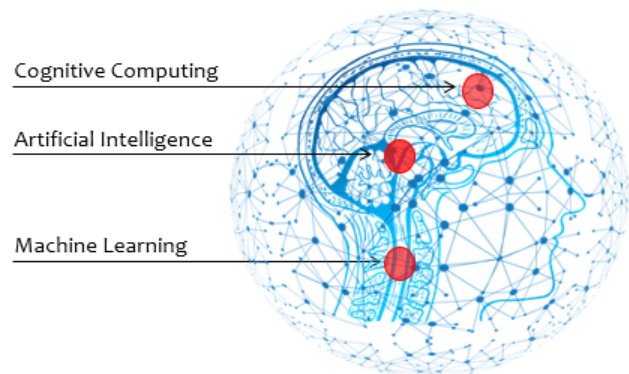

*Figure 6 An Analogy of Cognitive Computing: What is it?*

The term machine learning generally refers to the computerized techniques that basically take data, data streams and looking for patterns to quickly react based on what the algorithms are learning over time. It is analogous to the lower part of our brain stem, where if somebody touches something sharp, s/he immediately reacts to it because it's going quickly back and forth to the base of the brain. It is recognized as the lowest level of cognitive capability. Above that level, then, is artificial intelligence, which resembles the part of the human brain where humans start to develop a form of intelligence that begins to work together in a more complex way, using a more complex set of algorithms. Those are typical human functions for daily activities and similar functions of other animal creatures.

Cognition, however, is in the upper frontal lobe of the brain, as shown in the Figure 6. That is where humans differentiate, and that's where humans start to bring together different concepts and start reasoning. It not just about reading a simple image, speech, transferring a message from place to place; it rather brings together the human factors, contexts and all of the data sources into a more complex set of reasoning. And that's where cognitive is, where the exponential value can be unlocked in IoV.

Scientists and researchers in various disciplines, notably neuroscience, computer science, engineering and mathematics, use the term cognition to build and represent understanding by learning. In line with this, networking exclusively started using this term and applied it in the emerging Internet of Things (IoT) as Cognitive IoT (CIoT).

However, in the vehicular networking concept, Arooj et al. [15] define CIoV as a network-based framework for intelligent vehicles, where vehicles are primarily considered a context-aware agent. It is a sensory network of road elements such as vehicles and related infrastructure to receive real-time information from the road and surrounding physical world, including humans and the environment with minimal human interaction. It observes real data in live streams and conducts computing operations such as pattern matching using integrated intelligent algorithms in-vehicle applications. Thus evolved framework supports the communication between vehicles (V2V) and things (V2X) in the respective cloud and stores data. This heterogeneous datastore is used for knowledge discovery using advanced machine learning technologies such as deep learning, artificial neural networks and data mining techniques for better sensing, learning, prediction, and efficient resource management.

CIoV also applies the software-defined networking (SDN) and associated control services and systems such as Self- Organized Networking SON, Network Flow Virtualization (NFV), and network slicing to enable cloud-based service. It supports high manageability, high controllability, high operationalization and credibility of the network to increase driving assistance and improve road safety. The evolved C-IoV host multiple users, multiple vehicles, multiple things and multiple networks and initiates intelligence cooperation. It offers an in-depth integration of the human-vehicle-thing-environment with services and resources, increases transportation efficiency, improves the service level of cities, and ensures safe and comfortable travel [5, 15, 16, 18].

Therefore, it can be said that the emerging C-IoV is a paradigm that goes beyond connected vehicles. It supports storing, analyzing, and exchanging context-aware information among road transportation entities more intelligently.

*B.* C-IoV Framework

C-IoV is an advanced and well-defined framework evolved from IoV. In contrast to IoV, the CIoV offers hierarchical cognitive engines, i.e., control and cognition, and conducts joint analysis in both physical and network data space. From the network perspective, it can be said that VANET is a sub-network of IoV. And C-IoV is a refined and enhanced paradigm of IoV that offers more sophistication, management and security. Figure 7 shows the conceptual diagram of future transportation in relation to the smart city, outlining the cognitive engine within the framework of C-IoV.

The evolved cognition and control layer is part of the internet cloud that allows initiating cognition process utilizing cloud-based advanced features such as storing, processing, and computing the data collected from the lower layers of the CIOV architecture. The lower layers fundamentally consist of the communication layer and sensory, physical layer where a range of short to long-range communication technologies (cellular and non-cellular) and protocols resides. The sensory, physical layer addresses both in-vehicle and roadside heterogeneous sensors, ranging from wearables to camera to external GPS with data acquisition ability during the vehicle's movement [5, 10, 15, 19-21].

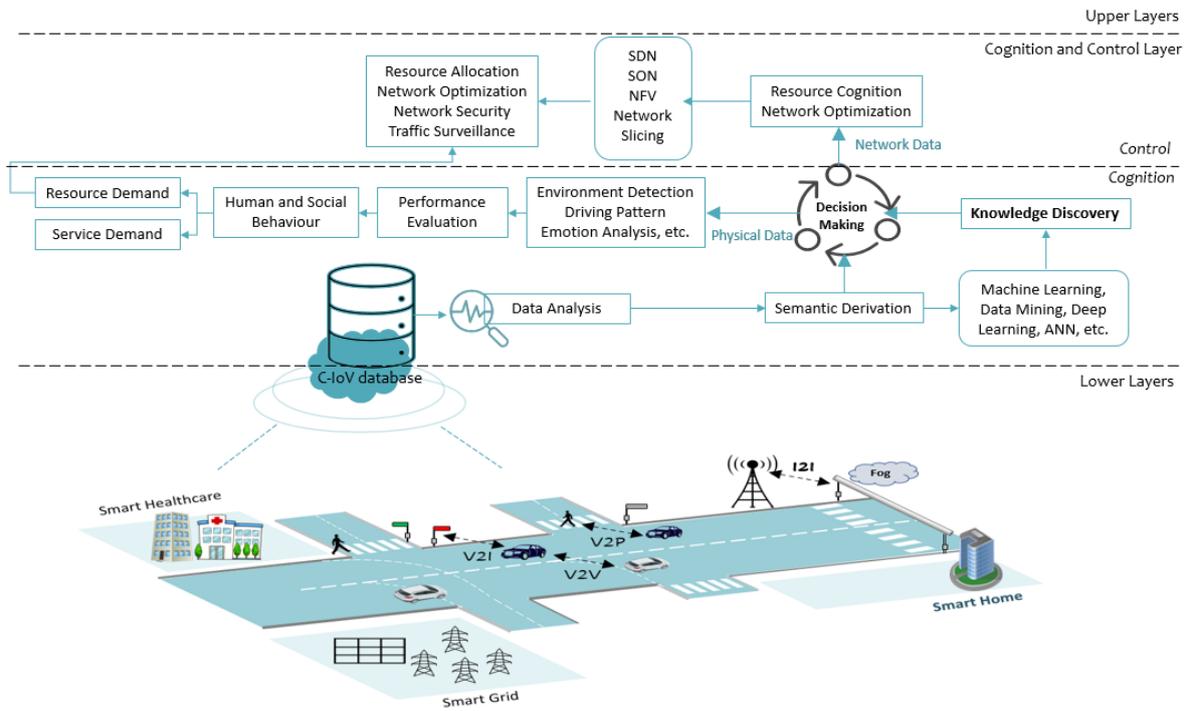

*Figure 7 Evolved Layer of Cognition and Control in C-IoV Framework*

The upper layer is the application layer responsible for management and service to coordinate and cooperate among different users and parties. It is also responsible for designing application services from intelligent transportation applications such as intelligent driving, intelligent transport management to customized application services such as driving guidance, emotion monitoring, etc [5, 16, 22].

The cognition engine slices between those layers serve as the bridge between the real-world observations and application realization. The fundamental purpose is to create knowledge from received data and treat them accordingly to the service required. In general, the physical data is intended to be processed to detect the environment, driving pattern, or drivers' or passengers' emotion analysis. On the other hand, resource cognition is derived from network data to conduct real-time traffic surveillance network optimization, enforcing security and resource allocation.

Overall, as shown in Figure 6, the key process involves data processing of the received data to understand appropriate use and management through several steps, e.g., semantic derivation and knowledge discovery, where both lead to the decision-making process to conduct behavioral analysis and system performance evolution.

*C.* Use case Analysis

Primarily the services enabled by the evolved C-IoV can be divided into two categories, strategic service and control service, as shown in Figure 8.

The purpose of the strategic service is to process and analyze data flow using cloud-based computing and storage technologies. The strategic services are enabled through different operations and analysis tools. Examples of some services include monitorization, behavior analysis, optimization of the network, pattern analysis and more. The indicated tools primarily are machine learning, neural networking, deep learning, distributed and federated learning and other artificial intelligence-based tools [4] [10].

On the other hand, the control services are responsible for determining system performances by utilizing a variety of different facilities such as network optimization, resource allocation, network-wide Security, etc. The technologies used to allow these services include software-defined networks, network slicing and more [4] [10]. Let us consider the following two scenarios to understand how road transportation under the evolved framework can enable travel safety.

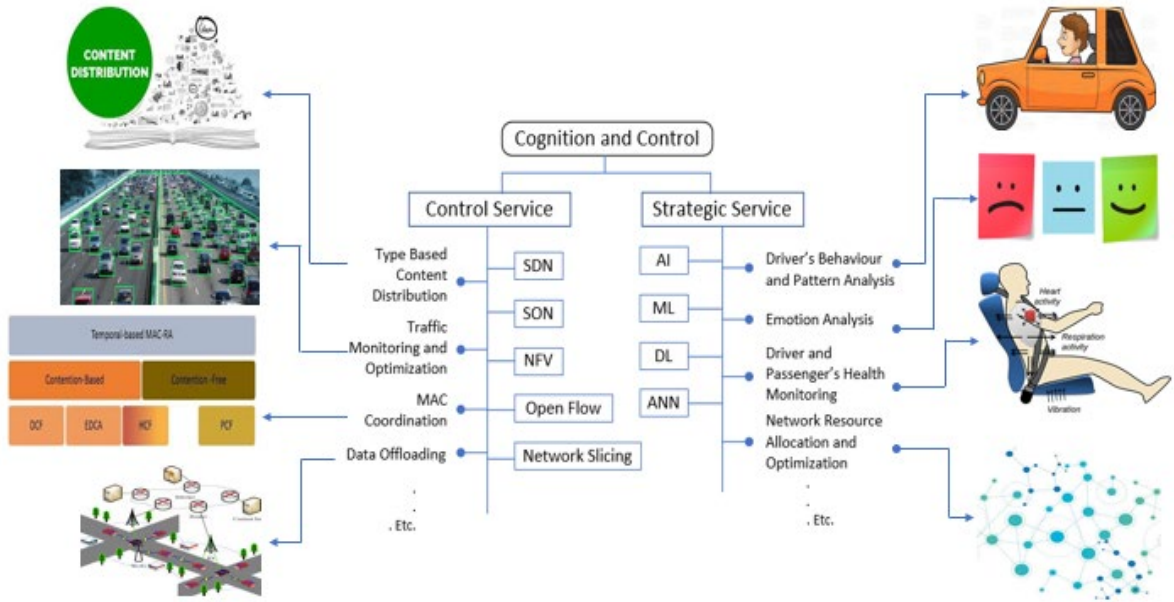

*Figure 8 Services of C IOV: Strategic Service and Control Service*

1) *Application Scenario-1*

Within the CIoV framework service, the smart vehicles are assumed to be capable of detecting, disseminating, and intelligently providing the content. Ideally, a myriad of sensors works together in the future look of the vehicular environment. Which includes intra-vehicle and inter-vehicle sensors for sensing and receiving information among road users, such as pedestrians, route environments, and other real-world contexts. To understand the typical application of CIoV, we can think of a situation where a traveler on a car is traveling home from work. Suddenly, assume that the vehicle senses some irregular seismic waves, raised temperature of the environment, high magnitude, and airflow frequency. These irregularities are detected on board the vehicle using various sensors. At the same time, vehicle continuously sharing the data with the cloud. Suppose other connected smart vehicles also observe the same abnormality from the same road and send the data to the cloud engine. At this point, the cognition engine will inherently analyze the data to detect any disaster that may occur proactively. It will initiate necessary measures at this stage, including the generation of early warnings and safe navigational route for the way home or safe place [15].

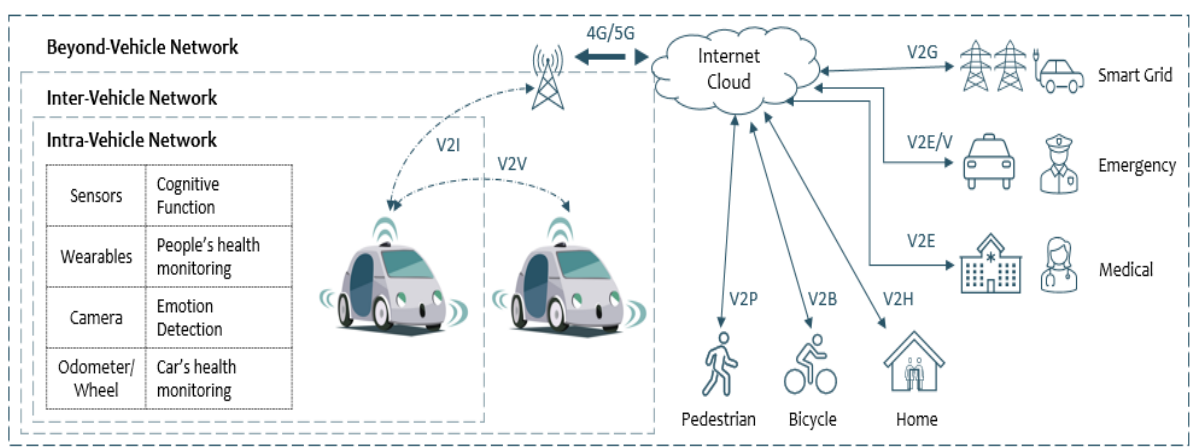

*Figure 9 Ubiquitous communication of Future Transportation System: Cognitive ITS [5]*

2) *Application Scenario 2:*

Let us consider another scenario where the future driving scenario can help the vehicle's driver or passenger get sick or unwell. Under the proposed CIoV framework, the deployed intra-vehicle network can carry out people's emotion detection and analysis. It can also perform driving behavior surveillance and physical health surveillance. Such surveillance is usually being conducted using onboard cameras and embedded sensors. For example, the intra-vehicle network camera can extract and analyze the driver and passengers' facial expressions. It can detect the eyelid state of the commuter to understand sleepy or dizzy behavior. Combining other data such as steering wheel movement for drivers, odometer readings can also be recorded to predict the subject's emotion or physical condition. Utilizing this information can deduce to generate necessary warnings to prevent any accident or take any measures needed as required. This level of operation can be achieved within the context of existing autonomous vehicular concepts through onboard computation. However, from the perspective of cognitive facilities, the evolved CIoV framework offers more resource support, such as the cellular network's communication resources, the remote data center's computing resources, nearby fog or edge devices such as roadside units or other vehicles. These resources can be used to conduct comprehensive condition analysis extensively for the sick passenger or driver. Simultaneously, the cognitive engine initiates some other necessary actions such as contacting the ambulance, doctor, and family members at home and sharing the onboard analysis result carried out using sensors to the doctor. All those activities would be taken autonomously without human involvement is the sole role of the evolved cognition engine that certainly can increase the survival of the sick on the road [5].

V. WHAT SECURITY, PRIVACY AND TRUST MEAN IN ITS?

Generally, in short, security in a cyber system refers to safeguarding data, privacy refers to safeguarding identity, and trust indicates a reliable relationship between parties. However, cybersecurity in practice is a set of technological measures ensuring that a system performs without interruption, achieving the tasks while mitigating unintended, external, unauthorized and unexpected interference. On the other hand, privacy can be defined as the reliance that the confidentiality of, and access to, sensitive information (such as Personal Identifiable Information (PII)) about the system or entity is protected and concealed. Privacy is often seen as an aspect of security, an affordance of confidentiality because a secure system should protect its users' privacy. Security also plays a central role in preventing service failures and cultivating trust in the network.

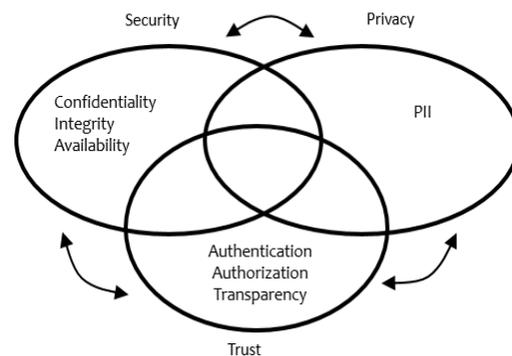

*Figure 10 Inter dependency between Security, Privacy and Trust in cyber physical system*

With the rapid advancement of ICT and the necessity of multifaceted connectivity, cybersecurity is becoming a challenging concern nowadays in every enterprise and industrial systems where transportation is no different. Direct attacks by malicious parties and security vulnerabilities expose the data of cyber technology-enabled transportation technologies. It is a threat to privacy and can be the reason for developing untrusted connections, which could jeopardize the entire system. Since transportation is a critical sector directly influencing humans, any technological failure can lead to road fatalities or property losses; in extreme cases, it can have catastrophic consequences, including death. Therefore, successful, safe and secure deployment of the intelligent transportation system is vitally important, which depends on the design of the cyber secure framework. While aiming for that, it is important to understand vulnerability and attack surface that may be associated with the evolved transportation. In the following subsections, at first, the high-level overview of the attack surface concerning the security and privacy of C-IoV is discussed. Later potential trust issues are presented.

A. *Attack Surface in evolved Intelligent Transporation System*

The evolved paradigm of transportation depicted in the above sections means, the entire system, both in automated and connected vehicular means, now carries more sensitive information. This includes vehicular telematics data to individual health data, which must be managed securely and privately. However, cyber threats are eminent as the transportation system is exposed due to the high accessibility and interdependence among systems that create new possibilities for different cyber-attacks and vulnerabilities.

In general, security breaches can happen through cyber system vulnerabilities that an attacker can exploit to enter the system. The attacker can also breach the system directly by other means, such as by injecting malware through hardware or communication channels or hardware ports. All are system entries termed as Attack Surface, shown in Figure 11 [23].

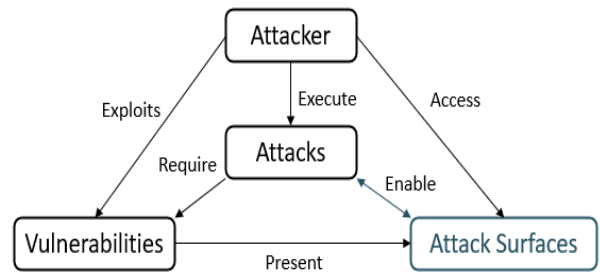

*Figure 11 Attack Surface: A relation between security-related terms in ITS.*

As shown in Figure 11, any system vulnerability leads to an attack surface that the Attackers can exploit to access the system. An attacker can also execute a series of small attacks to expose a larger attack surface to get into the system unauthorizedly. However, most often, an attacker requires an understanding of system vulnerabilities to execute an attack.

As the emerging ITS is exclusively a hybrid system over ubiquitous connections, the attack surface extends from hardware to service, software to applications. For instance, all the automated car's hardware systems are attack surface. When vehicles communicate with each other, any infected vehicle can be a potential source to breach the security.

The attack surface can be the onboard system interface or a remote interface. Any physical interface, such as Onboard hardware like the On-Board Diagnostics (OBD) port, allows the vehicle to access the roadside infrastructure, disk reader, USB port, etc. can be the sources for the attack to gain access to the vehicle. Other physical parts of the bodies from different OEMs are also potential interfaces and are considered an attack surface. Compared with that, the remote attack surface is all possible entities that the vehicles are wirelessly connected. Several wireless technologies, such as DSRC, 4G/5G, Bluetooth, Wi-Fi, Remote Keyless Entry, GNSS, etc., can be the remote surface access interface.

Furthermore, the vehicles rely on an external cloud, edge, and fog infrastructure for processing and storing of data. The remote infrastructure that handles the vehicles' data must not be vulnerable, meaning it should not suffer from any hardware or software faults and failures. Faulty software may expose an attacking surface to malicious attackers, who may then intercept the vehicles' private and sensitive data and use it for malicious purposes. In addition to the data leakage, a malicious attacker may also compromise the integrity of the computation processes on the remote node, which would lead to either incorrect or delayed output. Making critical decisions relies on a remote infrastructure is not safe for the vehicles. Hence, the attack surface on the remote infrastructure such as cloud must also be minimized.

Overall, the attack surface resides within the vehicle, among the vehicle's connections and beyond the vehicle interactions. In a top-layer view, the attack surfaces for all potential cyber-attacks and data breaches fundamentally can be viewed again in a three-tier diagram shown in Figure 12, where intra vehicle-based ECUs, between vehicles communication and cloud, cloud-connected parties are seen as the potential attack surface.

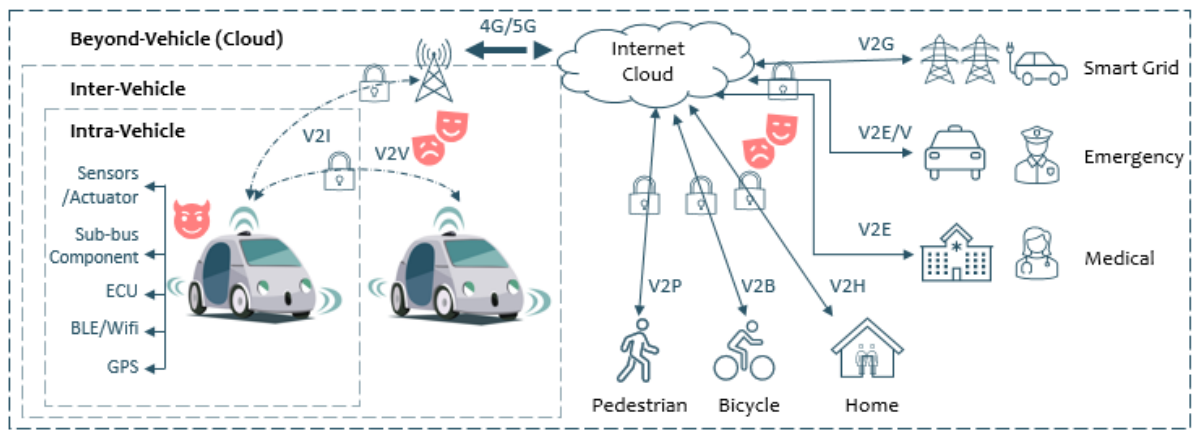

*Figure 62 Three-Tire view of the possible attack surface*

### B. Trust in the evolved C-ITS framework

Trust has many meanings across different disciplines, however, trust in networking implies a reliable multiparty relationship. It narrates the state to which a network node or entity accepts others' dependence [24]. Therefore, any network party has to provide a notion about the condition such as the correctness of the data within the context. In a vehicular network, this notion or signature is complex and depends on accuracy, timeliness, accessibility, and also interoperability. And with the introduction of large volumes of data through C-IoV, trust becomes diversified.

The multifaced and growing connectivity within and beyond the smart city's transportation system, trust becomes more important than ever. The evolved C-IoV in the smart city context is in a process to connect the wide real physical world with digital control by sensing the world, understanding it and programming it. This leads to the possibility of losing information and the loss of control over the device or host. Any breach of information security can endanger commuters' lives on the road and social loss in personal life. To achieve absolute security to the emerging transportation system and exchange reliable and authentic information among network nodes, peers, and the cloud, an attack-free and trusted environment are foremost.

Within the vehicular network context, trust ensures expected result of communicating with peers in every session. However, the concern is that the outcomes can be a positive value or being hacked and or cheated somehow. When processing application data on remote infrastructure, including cloud, edge, and fog nodes, it is important to ensure that the remote nodes are not compromised and they are strictly adhering to the service-usage agreements. However, a compromised node may behave maliciously. Imagine a vehicle is employing an object detection application to count the objects in a long distance. Due to the lack of local processing power and soft real-time requirements, the vehicle offloads the images and the object detection code to a remote node. If a remote node is honest and uncompromised, it would correctly execute the code using the vehicle's provided input image and then return the number of objects found in the image. However, when a remote node is compromised, it may simply return random results without running any computation, which may hamper the performance and the responsiveness of the object detection application. Note that this attack may lead to catastrophic consequences if it is a safety-critical application.

In the evolved framework of C-IoV, there exist multilevel interactions. For example, road-level interaction where vehicles wirelessly communicate with other vehicles, sensors, pedestrians, roadside units, intelligent city entities. Interaction and cooperation among multiple cloud service providers (CSPs) as a single cloud would not provide all devices with satisfactory quality of services due to performance and the high cost of cloud deployment or country policy. Additionally, while coordinating with other smart city services, the transportation cloud would have to interact with the targeted service cloud to enable the services shown in Figure 13. Ensuring trust among all these untrusted parties is a highly desirable challenge of such a network.

However, with such multifaceted interaction, CIoV is exposed to a range of threats to security and privacy. Any dishonest and misbehaving peers and malicious CSP in the system is a major concern, increasing the system's vulnerability and endangering lives. Ensuring trust among all these untrusted parties is a highly desirable challenge of such a network.

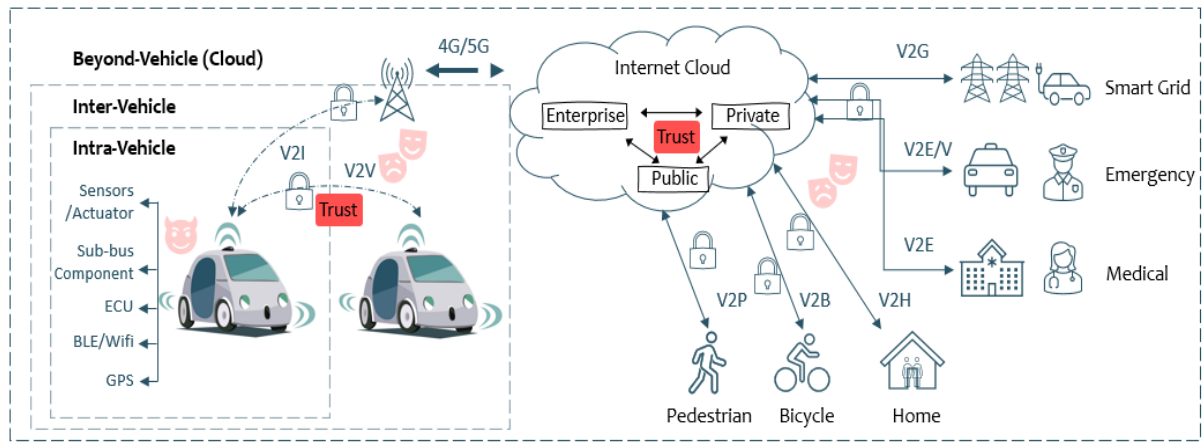

*Figure 13 Issue of Trust in Smart Transporation*

VI. PROSPECTIVE COUNTERMEASURES ENABLED BY COGNITIVE ENGINE IN LEGACY SECURITY ISSUES

The evolved transportation framework can face significant security issues over the predecessors due to its multifaced connectivity with other sectors towards realizing the smart city, which would introduce additional attack surfaces to the system. However, the technology associated with the evolved cognitive engine in the emerging C-IoV framework can improve the system's overall security. The IoV framework suffers many security issues that primarily stem from the network side of IoV [25]. In relation to that, we have identified four areas where cognitive technology, as it is defined, can be used to improve the existing security issues of IoV [26-28], are following:

*a)* Authentication: Authentication, in general, allows network devices to distinguish between outside attacks and source nodes. It is an essential security feature for networks that address identity-based attacks such as spoofing and Sybil attacks [29]. Ideally, the authentication process starts by comparing the physical layer characteristics to the transmitter's characteristics following in determining the authentication of the transmission. This can be done by comparing the characteristics to a particular threshold built on the classification model's accuracy. Over time, a set of machine learning-based techniques, for example, reinforce learning techniques such as Q-Learning, Dyna-Q, and deep Q-network (DQN) has been proved to select the required threshold to acquire authentication accuracy [30-32]. Besides, Convolution Neural Network (CNN) model shows promise to extract driver behavioral characteristics from the original data in the experimental vehicles achieving higher accuracy driver identification that can be potentially used for Authentication [33].

*b)* Access control and verification: The use of access control prevents unauthorized users from accessing resources within the network. The emerging transportation network is essentially heterogeneous, and it is challenging to design an effective access control mechanism for such a network. However, some machine learning-based access control mechanisms such as Support Vector Machines (SVMs), K-Nearest Neighbors (K-NNs), and Nearest Neighbors (NNs) have come up with optimal solutions that can be employed to achieve access control support from and within C-IoV [32, 34].

Within the evolved paradigm, it is possible to integrate a verifiable computation framework that allows the vehicle or other third parties to verify the correctness of data independently. Emerging cryptographic implementations such as ZKSnarks are offering viable solutions. Besides, a new class of verifiable computation systems based on run-time profiling is being introduced – verifiable Python (vPython) is a new verifiable computation framework that helps the application developers to gather proof from a remote node that it has correctly executed the computation [44].

*c)* Secure network offloading: Secure offloading allows for networked devices in the vehicular network to use external, storage, and cloud computation resources for specific tasks that would require heavy computational power. Q-Learning, a machine learning tool and a feature of the proposed cognitive engine, can be used to identify an optimal threshold for authentication, can also be used to identify the best rate of offloading data to provide security against specific attacks, such as jamming and spoofing attacks.

*d)* Malware detection: Cloud-based vehicular network is prone to attack from a wide range of malware such as viruses, worms, and trojans [35]. Machine learning-based malware detection techniques, for example, supervised learning techniques such as naïve Bayes, neural networks, or K-nearest neighbor, can be used to label the traffic going into the IoV network and build classification models to detect network intrusion [32].

Along with the above-mentioned support, the cognitive engine can be used for continuous software updates. Contemporary software frameworks and operating systems are prone to bugs. The software community is continuously describing the potential vulnerabilities and release patches to minimize the damage. Therefore, running outdated software frameworks and operating systems would open up a large attack surface, which can be minimized by employing good software and security practices. Again, when relying on third parties for computation and storage, it is important to verify their actions to ensure correctness and safety. By including verifiable computation and storage solutions within the concept of cognitive framework, the application developers can get assurance from the remote infrastructures. Additionally, the hierarchical cloud model with a robust centralized cognition model can fit the emerging federated learning model to be adopted. In federated learning, each independent vehicle need not share their complete data to a central server for learning purposes. Instead, each vehicle can run a learning algorithm locally within their vehicle and share only the updated model parameters with the remote server. This model helps the vehicle retain their sensitive data while still contributing to the learning process and contribute to security and privacy improvement.

Finally, the cloud-based C-IoV framework essentially gives a centralized infrastructure which is one of the inherent concerns as relying on a centralized architecture would lead to a central point of failure. A malicious attacker may compromise the infrastructure that serves the vehicles. Since the vehicular network is dispersed and inherently decentralized, decentralized frameworks involving blockchain and distributed ledger technologies help the application developers include stakeholders in the application processes, which prevents the central points of failure and provides more transparency. A cognitive engine can play a vital role in enabling such a hierarchy to apply decentralized solutions under centralization.

## VII. SECURITY AND PRIVACY CONCERNS IN C-IOV

Under the umbrella of C-IoV, a composite network can be observed; thus, system vulnerabilities can be a common reason to execute successful attacks. Besides, any active attacking mechanism such as injecting malware or even sufficient knowledge about the system can help attackers hack into a targeted system. It happens that vulnerabilities in the transportation system may exist; for example, simply it can be due to the weakness of the system design or software bugs that can be exploited to enter the system. Therefore, vulnerabilities are subjected to exist at all tiers, intra, inter, and beyond vehicular networks.

In the case of autonomous intra-vehicular networks, many recent studies highlighted the internal bus's limitation and weakness, which possibly can allow unauthorized access to the system without any restriction [21, 36-39]. For inter-vehicular networks, many vulnerabilities around DSRC (IEEE 802.11p) have also been addressed; for example, Ucar et al. [40] conducted a vulnerability test that shows the technology gap using omnidirectional antennas that is a potential cause of jamming attacks. Similarly, Lyamin et al. [41] show jamming DoS attacks in DSRC while exchanged beacons are corrupted. The recent proposal of IEEE 802.11 -OCB as a replacement of IEEE 802.11p does not offer cryptographic protection since it operates outside the context of a basic service set. This also indicates the potential insecurity, which needs to investigate [42]. Research [43, 44] shows the major vulnerabilities in cellular networks that lead to IP-based attacks, eavesdropping, spoofing, DDoS attacks, and many other well-known, well studied. Global Navigation Satellite System (GNSS), as an integral part of the inter-vehicular communication system, is also subjected to exhibit system, propagation, and interference-related vulnerabilities, leading to Jamming the service Spoofing the network [45-48]. Similarly, cloud-based beyond vehicular networks is prone to have many vulnerabilities that are a potential issue for the emerging ITS.

Besides, the vehicle is supposed to generate massive data where some are sensitive, such as personal health-related data, vehicle registration, condition or location data, or data about the nearby peer vehicle or cloud. Therefore, these are serious privacy issues that need to pay attention to dealing with identity privacy and location privacy. However, this presentation's scope covers the security supports and related issues with the evolved cognitive engine inside the cloud.

The following section presents some security concerns of the evolved cognitive engine in the cloud of the evolved C-IoV framework. In this short presentation, we have identified the fundamental security issues in relation to the cloud and the machine learning algorithms with the cognitive engine that may cause security concerns for the proposed paradigm.

### A. Issues related to the cloud computing of the cognitive engine

Cloud computing has many unique security concerns that relate heavily to the technologies used within the cognitive engine. The engine relies on the cloud to store the data to allow vehicles to learn patterns and much more; security concerns already present in cloud computing have been considered to IoV. Some high-level security concerns for the cognitive engine on the base of cloud computing/storage are listed as following [49-51]:

1. *Transborder data flow/data proliferation*: one of the significant issues that CIoV could potentially face with its users is that specific companies can access the stored data with or without permission from the user. This makes the issue of data integrity and privacy a severe and genuine concern for users.

2. *Access*: since the data is always stored to allow for training, as well as many other uses for cognitive technology within CIoV, there is the prospect that users' data could be compromised by attackers breaking into the cloud storage that keeps their data.

3. *Multitenancy*: Clouds run on multiple different machines; this leaves the integrity of the data of CIoV at risk. This leads the cloud to be more susceptible to attacks on the cloud infrastructure.

4. *Trust*: As the CIoV model stores substantial data relating to the users' information from many different devices, e.g., mobile phones and cameras, users have a lack of total trust as there is a significant fear of the systems collapsing and, therefore, the possibility of losing data.

### B. Issues related to learning algorithms in the cognitive engine

The proposed cognitive engine in the CIoV framework impacts the traffic industry in a new and automotive way. However, security breaches into the learning algorithm can negatively impact the users' private data. Security issues related to the machine/deep learning algorithm can make the classification model misinterpret specific data or be taught incorrect data.

There are three classes of concerns with the use of a machine learning algorithm that is needed to be taken into account in realizing effectivity of the cognitive engine;

a. Influence attacks can influence the classifier model by altering or disrupting the classification phase.

b. Security violation is a form of attack aimed at providing false negatives that allow hostile input to a system or possible availability, which focuses on false positives that deny benign input from accessing the system.

c. Specificity, which focuses on allowing a specific intrusion/disruption. This attack also can cause general mayhem in the training model [52].

Three significant security concerns related to machine learning have been identified [27, 52-54]. These three security concerns consist of:

1. *Evasion attacks*: This is a form of attack on machine learning that allows the attacker to manipulate malicious samples with the intent to evade detection. This attack is hazardous to any technology using machine learning that connects to the internet because it is designed to bypass classifiers, allowing attackers to sneak into the network without detection.

2. *Data Poisoning*: Data poisoning is very different from evasion as instead of avoiding detection, poisoning is about altering the training data for machine learning. This is done by feeding polluted training data into the classifier model, giving rise to what is classified as true and false regarding the attackers' favour. The most common form of poisoning is done by switching the classifiers' understanding of what good inputs are and what are bad inputs. This could potentially be one of the more worrying security concerns for CIoV as many of the decisions that the CIoV engine will make can be potentially flipped to have the worst outcomes [55] (e.g. V2V meter distance of safety).

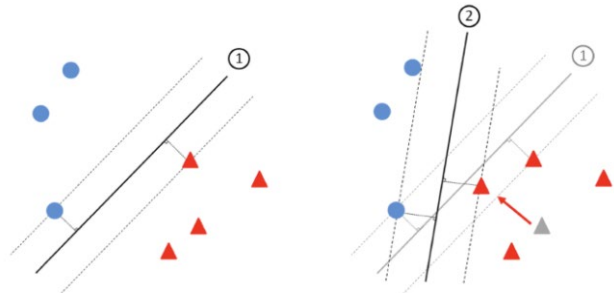

*Figure 14 Classifier decision boundary from the default (left) being altered from poisoning, causing a massive change in decision choices (right).*

1. *Model stealing techniques*: This security issue is one of the least likely to happen but can still pose a serious concern. This concern is more about recovering models/information that was used for training with the specific use of the training data. This is about how the training models represent intellectual information that has been used to train sensitive data. These forms of data could include anything on users' mobile phones, driving locations as well as medical conditions and more.

Issues span further with machine/deep learning for CIoV when it comes to the data learning algorithm. Because CIoV is proposed to be one of the recently most advanced and widespread applications of AI, there is a high chance of misinterpretation of the training data without the need of an attacker. The learning patterns in machine/deep learning can be brittle as the model works on data that is like natural data. If unique data that is slightly different from the other training data is used on the learning algorithm, this can cause the model to fail completely. This can bring the theory about how an attacker does not necessarily need to use cyber-attacks to affect CIoV's AI system [56] negatively. For example, one could cover a 'Stop' sign with some tape, and the AI system would not have a clue on how to react; in fact, it might merely resume course and not stop at all.

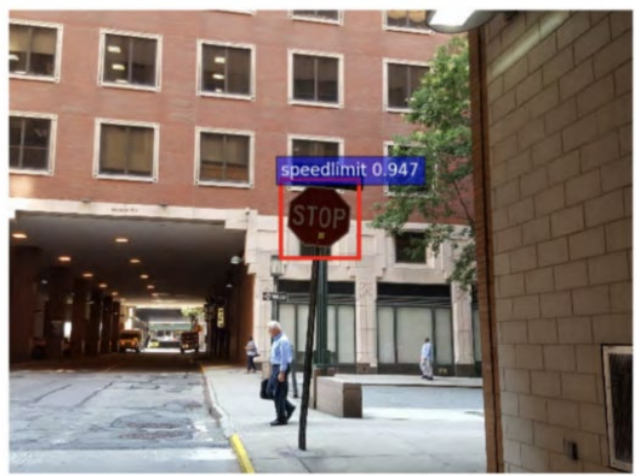

*Figure 15 Machine learning algorithm misses classifying 'Stop" sign taken from [9]*

## VIII. CONCLUSION

The advent of artificial intelligence and cloud computing has brought some new prospects for the transportation system's future realization. Simultaneously, new technology faces unique challenges due to the heterogenic structure and critical nature of the transportation system and its application. This chapter presents an overview of the future transportation expectation. Systematically it illustrates the evolution of Intelligent Transportation System, including the recent development of the smart city-centric framework Cognitive Internet of Vehicles. Later it covers the Security, Privacy, and Trust issues related to the emerging intelligent transportation system. More specifically, it discussed the enabling security supports by the evolved framework and the potential attack surface that extends due to the multifarious communication. Overall, the chapter deduced that CIoV has the potential to meet great expectations by understanding that the technologies involved with the new integrated Cognitive engine will provide significant benefits that can reduce human errors and fatal crashes. This chapter indicates that it has potential for much wide use of cognitive engine technology and therefore further need to explore.